\pdfoutput=1
\documentclass[a4paper,fleqn,usenatbib]{mnras}

\usepackage[T1]{fontenc}
\usepackage{ae,aecompl}

\usepackage{graphicx}	% Including figure files
\usepackage{amsmath}	% Advanced maths commands
\usepackage{amssymb}	% Extra maths symbols

\title[The quest for Neptune's  H$_3^+$]{The quest for H$_3^+$ at Neptune: deep burn observations with NASA IRTF iSHELL}

\author[H. Melin et al.]{
Henrik Melin,$^{1}$\thanks{E-mail: henrik.melin@leicester.ac.uk (HM)}
L. N. Fletcher,$^{1}$
T. S. Stallard, $^{1}$ 
R. E. Johnson, $^{1}$ \newauthor
J. O'Donoghue, $^{2}$
L. Moore $^{3}$
and P. T. Donnelly$^{1}$
\\
$^{1}$Department of Physics \& Astronomy, University of Leicester, Leicester, UK \\
$^{2}$Planetary Magnetospheres Laboratory, NASA Goddard Space Flight Center, Greenbelt, Maryland, USA\\
$^{3}$Center for Space Physics, Boston University, Massachusetts, USA
}

\date{Accepted XXX. Received YYY; in original form ZZZ}

\pubyear{2015}

\begin{document}
\label{firstpage}
\pagerange{\pageref{firstpage}--\pageref{lastpage}}
\maketitle

\begin{abstract}
Emission from the molecular ion H$_3^+$ is a powerful diagnostic of the upper atmosphere of Jupiter, Saturn, and Uranus, but it remains undetected at Neptune. In search of this emission, we present near-infrared spectral observations of Neptune between 3.93 and 4.00 $\mu$m taken with the newly commissioned iSHELL instrument on the NASA Infrared Telescope Facility in Hawaii, obtained  17-20 August 2017. We spent 15.4 h integrating across the disk of the planet, yet were unable to unambiguously identify any H$_3^+$ line emissions. Assuming a temperature of 550 K, we derive an upper limit on the column integrated density of $1.0^{+1.2}_{-0.8}\times10^{13}$ m$^{-2}$, which is an improvement of 30\% on the best previous observational constraint. This result means that models are over-estimating the density by at least a factor of 5, highlighting the need for renewed modelling efforts. A potential solution is strong vertical mixing of polyatomic neutral species from Neptune's upper stratosphere to the thermosphere, reacting with H$_3^+$, thus greatly reducing the column integrated H$_3^+$ densities. This upper limit also provide constraints on future attempts at detecting H$_3^+$ using the James Webb Space Telescope. 
\end{abstract}


%%%%%%%%%%%%%%%%%%%%%%%%%%%%%%%%%%%%%%%%%%%%%%%%%%

%%%%%%%%%%%%%%%%% BODY OF PAPER %%%%%%%%%%%%%%%%%%

\section{Introduction}

Neptune is the most distant planet in our solar system orbiting the Sun at a distance of 39 AU, receiving only 2\% of the solar flux that Jupiter receives. It has only been visited by a single spacecraft, Voyager 2, which flew by in August 1989. It has a magnetic field tilted from the rotational axis by 47$^{\circ}$ \citep{1989Sci...246.1473N}, which positions the magnetic poles at mid-latitudes. Therefore, unlike at Jupiter and Saturn, we do not expect auroral emissions to be located close to the rotational poles. In addition, the magnetic dipole is offset from the centre of the planet by 0.55 R$_N$ (Neptune radii, 24,622 km) towards the southern hemisphere, rendering the northern auroral region large and the southern small, reminiscent of Uranus \citep{2009JGRA..11411206H}. Disk-wide ultraviolet H$_2$ emissions were observed at Neptune by Voyager 2 on the nightside \citep{1990GeoRL..17.1693S}, with the addition of a brightness peak in the southern hemisphere that was interpreted as auroral emission about the southern magnetic pole. This tentative result remains the only observation of the aurora of Neptune. Modelling by \cite{2015JGRA..120..479M} suggests that reconnection at the magnetospause is less favoured at Neptune than at Jupiter, Saturn, or Uranus, making the likelihood of a bright aurora small. 

The molecular ion H$_3^+$ is produced in the upper atmosphere of a giant planet via this exothermic reaction:
\begin{equation}
H_2^+ + H_2 \rightarrow H_3^+ + H
\end{equation}
where the H$_2^+$ reactant is ionised by either extreme ultraviolet (EUV) solar photons, or by  charged auroral particles impacting the atmosphere. Observations of H$_3^+$ have played a critical role in advancing our understanding of the ionosphere, upper atmosphere, aurora, and magnetic field at Jupiter, Saturn, and Uranus \citep[e.g.][]{2004jpsm.book..185Y, miller_2010, 2011ApJ...729..134M}. We can derive upper-atmospheric temperatures and densities by observing the spectrum of H$_3^+$, in addition to providing a measure of energy lost via radiation to space. The morphology of the observed H$_3^+$ emission produced by auroral processes is directly related to where in the magnetosphere charged particles originate, revealing the planet's magnetic configuration.

\begin{table}
\begin{centering}
\begin{tabular}{ l l l l }
\hline
Observing Date & Neptune (h) & Uranus (h) & Seeing ($^{\prime\prime}$) \\
\hline
\hline
2017 August 17 & 3.7 & 0.2 & 0.6 \\
2017 August 18 & 3.7 & 0.4 & 0.6 \\
2017 August 19 & 3.8 & - & 0.7 \\
2017 August 20 & 4.2 & 0.5 & 0.6 \\
2017 August 21& - & - & -  \\
\hline
\hline
TOTAL & 15.4 & 1.1 &  \\
\hline
\end{tabular}
\caption{The NASA IRTF iSHELL observations of Neptune and Uranus analysed in this study, with the times being the on-target integration-times, without overheads. We were awarded five second half nights in August 2017 (programme 2017B077). The first four nights were clear, but we were unable to observe on 21 August due to a snowstorm at the summit. \label{observations}}
\end{centering}
\end{table}

Using Voyager 2 radio occultations \cite{1995Sci...267..648L} derived an altitude profile of ionospheric electrons, from which they modelled the vertical distributions of several related species, including H$_3^+$. The ion had a peak density at about 1400 km ($\sim 1.1$ nbar) above the 1 bar level with a volumetric density of $1.1 \times 10^6$ m$^{-3}$. At this altitude, the temperature of the upper atmosphere is $\sim$550 K \citep{1989Sci...246.1459B}, although this altitude region coincides with the positive thermospheric temperature gradient, suggestive of significant uncertainties. The exospheric temperature derived from these occultations is $\sim$750 K \citep{1989Sci...246.1459B}. 

H$_3^+$ emissions from the giant planets are observed to be variable on both short and long time-scales \citep[e.g.][]{2013Icar..223..741M}. Short-term variability is driven mostly by the highly dynamic auroral process, producing a rapid changes in the ionisation rate of molecular hydrogen. These changes in auroral activity are caused by variable solar-wind conditions, or by changes in the production rate of plasma inside the planet's magnetosphere. Long-term changes are somewhat harder to elucidate, but are likely driven by changes in the ionisation rate of molecular hydrogen by changing solar conditions as we move through the solar cycle. These variabilities fluctuate around a baseline of average solar and auroral ionisation conditions. Via the analysis of H$_3^+$ spectra, we can investigate the physical mechanisms responsible for these changes. However, at Neptune, where H$_3^+$ remains undetected, we do not know the extent of any variability, or what the baseline level is.

Despite several attempts at detecting H$_3^+$ emissions from Neptune, none has been successful \citep{1993ApJ...405..761T, 2000AA...358L..83E, 2003AA...403L...7F, 2011MNRAS.410..641M}. Most recently, \cite{2011MNRAS.410..641M} used about 1 hour of Keck NIRSPEC data from 2006 \citep{1998SPIE.3354..566M} observations to derive an upper limit of the H$_3^+$ column density of $\sim1.5 \times 10^{13}$ m$^{-2}$ at an assumed temperature of 550 K. 

The iSHELL instrument at the NASA Infrared Telescope Facility, installed on the telescope in August 2016, combines high sensitivity with high spectral resolution, and provides our best present opportunity for long integrations to search for H$_3^+$ emissions at Neptune. Section \ref{secobs} describes the observations and instrument configuration, with Section \ref{secanalysis} describing the analysis of the data. In Section \ref{secresults} we discuss the results of the analysis and we end with conclusions in Section \ref{secconc}.

\begin{figure}
	\centering
	\includegraphics[width=\columnwidth]{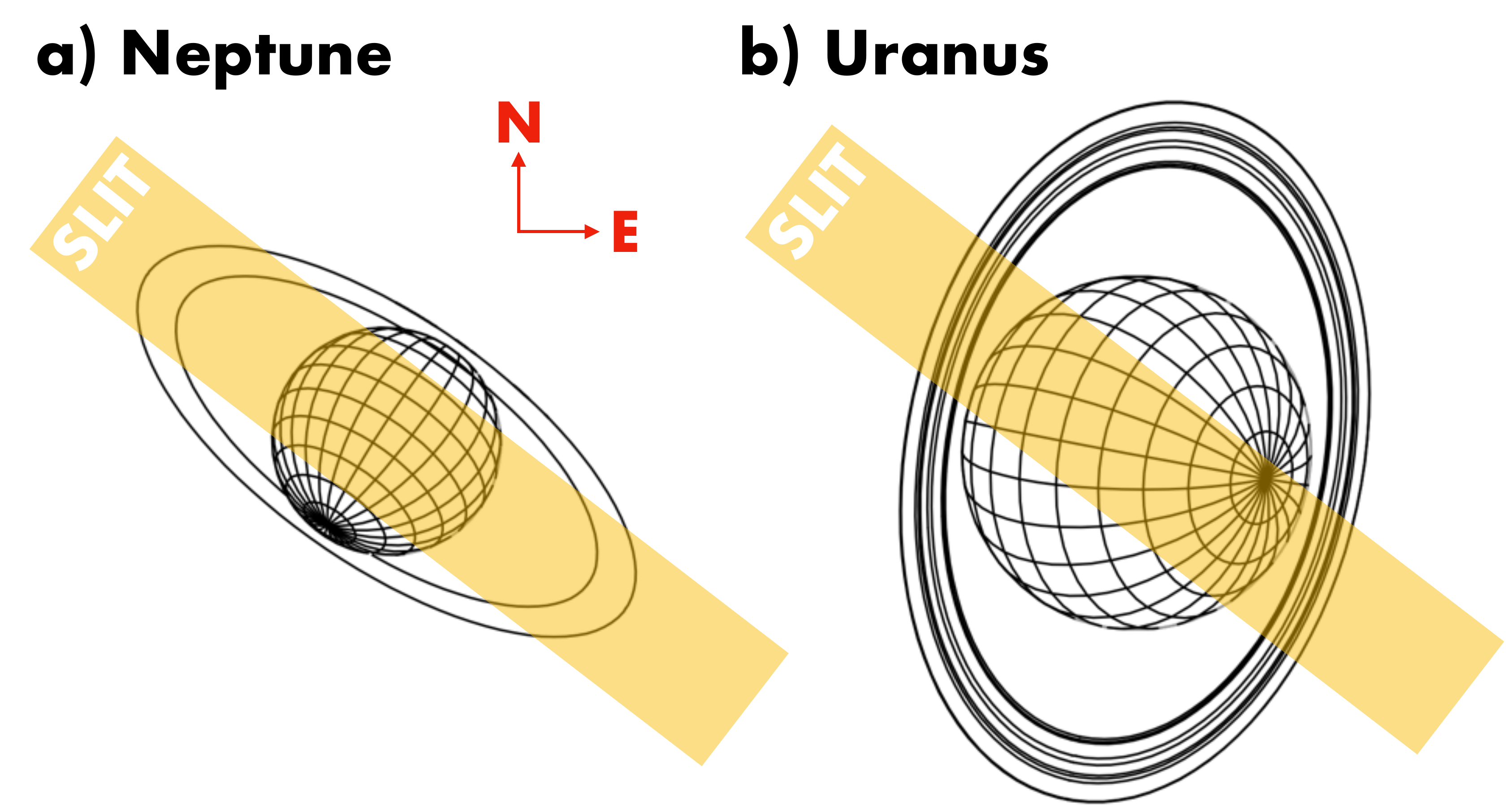}
	\caption{The observing geometry for the NASA IRTF iSHELL observations of Neptune and Uranus in August 2017. The 1.5$^{\prime\prime}$ wide and 15$^{\prime\prime}$ long slit was aligned along the planetographic equator of Neptune. The full length of the slit is not shown in this schematic. Neptune subtended 2.4$^{\prime\prime}$ in the sky, and Uranus subtended 3.6$^{\prime\prime}$. \label{geometry}}
\end{figure}

\begin{figure*}
	\centering
	\includegraphics[width=7in]{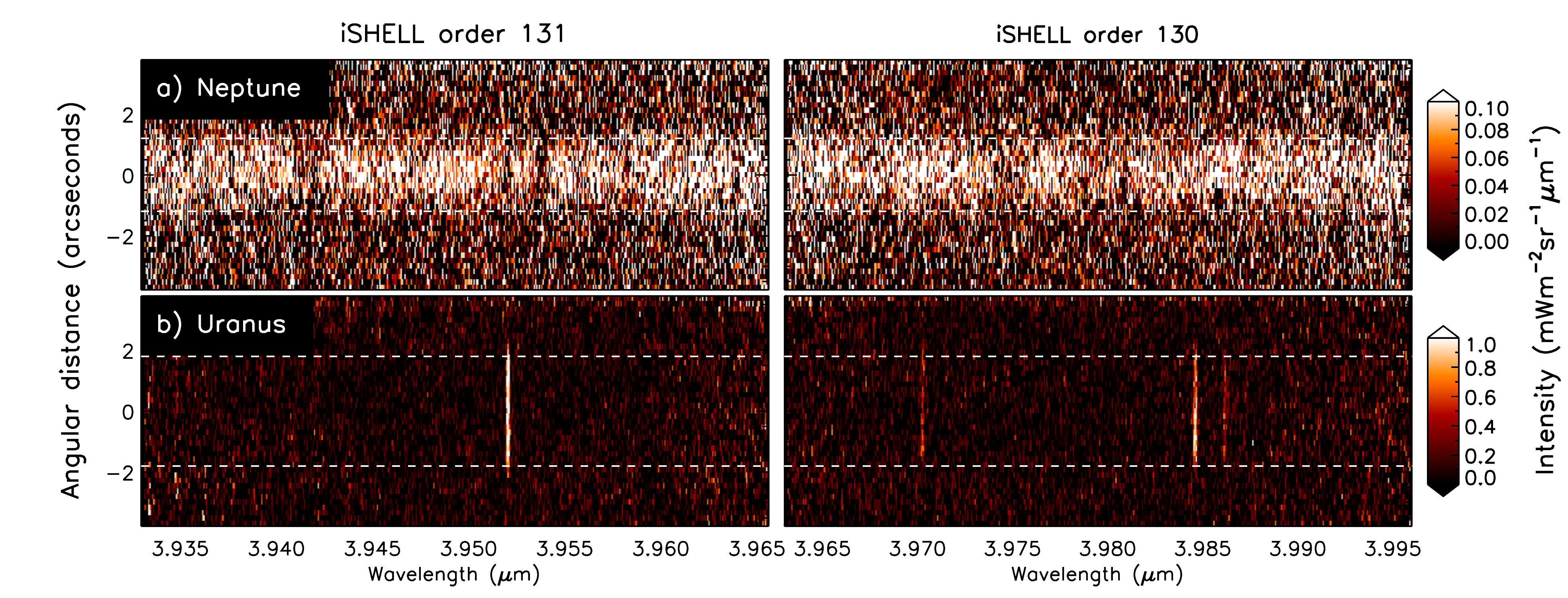}
	\caption{The medianed spectral images of Neptune and Uranus obtained in August 2017 using NASA IRTF iSHELL. The horizontal axis is the spectral dimension, containing over 4000 pixels, and the vertical is the spatial dimension. The dashed lines shows the angular size of each planet. Neptune has a distinct continuum emission, whilst Uranus has discreet H$_3^+$ emission lines.  \label{specims}}
\end{figure*}

\section{Observations} \label{secobs}

We used the NASA IRTF iSHELL spectrograph \citep{2016SPIE.9908E..84R} to observe Neptune and Uranus between 17 and 20 of August 2017 (UT). The observations are outlined in Table \ref{observations}. We used the 1.5$^{\prime\prime}$ wide slit covering most of the 2.4$^{\prime\prime}$ disk of Neptune, aligning the slit along the planetographic equator, with a position angle of 55$^{\circ}$ West of North. This slit produces a resolving power of $R = \Delta \lambda / \lambda \sim$17,500. The observing geometry can be seen in Figure \ref{geometry}. The slit length was 15$^{\prime\prime}$ allowing us to nod the telescope so that Neptune is on the slit in both the object (A) and sky (B) exposures. Each exposure was 120 s long, divided into two co-adds. We guided the telescope on Neptune, using the on-axis guider (Kyle) with the J-band filter (1.05 to 1.45 $\mu$m). 

% filter to guide the telescope trailed on Neptune, using the on-axis guider. The guiding was stable to within $\sim$0.5$^{\prime\prime}$. 

The Lp3 (L$^\prime$) iSHELL cross-disperser setting produces 13 spectral orders that are projected onto the infrared detector array, which has 2048 by 2048 pixels. These cover a wavelength range between 3.83 and 4.19 $\mu$m, with some overlap between each order. The angular projection of each spatial pixel is 0.18$^{\prime\prime}$ (or 0.08 R$_N$). We used our newly developed IDL iSHELL pipeline to reduce the observations, including standard flat-fielding and sky subtractions. The orders were straightened in both the vertical (spatial) and horizontal (spectral) dimensions, and wavelength calibrated using the telluric skylines. The spectra were flux calibrated using observations of the A0 star HR 830 (K magnitude 5.909). We only extracted the two orders that contained the H$_3^+$ Q(1, 0$^-$) and the Q(3, 0$^-$) emissions (orders 131 and 130 respectively), with each extracted spectral order having a dimension of 2048 by 110 pixels and a wavelength resolution of 0.016 nm. The spectral range of these two orders is 3.93 to 4.00 $\mu$m.

Our observing strategy was as follows: for each of the four nights that we observed (Table \ref{observations}), about 5 hours were spent integrating on Neptune, until an airmass of 2 was reached. Over the entire programme, we integrated on Neptune for 15.4 hours. We then performed short reference observations of Uranus, so that the location of the H$_3^+$ emission lines on the detector array could be experimentally determined, and our wavelength calibration validated. We spent a total of 1.1 hours integrating on Uranus.  The angular diameter of Uranus was 3.6$^{\prime\prime}$, and we used the same slit rotation as used for the Neptune observations -- see Figure \ref{geometry}. At the end of the night the flux calibration star was observed, in addition to flat-fields and other calibration frames.

The rotation period of Neptune is $16.11\pm0.05$ h \citep{1989Sci...246.1498W}, which means that for each Earth day, Neptune has completed $\sim$1.5 rotations, and we are faced with a 180$^\circ$ longitude phase difference on consecutive observing nights. Over four half nights of observing, all longitudes of Neptune will have been observed, albeit some at large viewing angles. Regardless, as will become clear, we lack the signal-to-noise to analyse any shorter components than the average of all observations. 

During these observations, Uranus was moving towards the Earth with a speed\footnote{Sourced from NASA Horizons at https://ssd.jpl.nasa.gov/horizons.cgi} of 26 kms$^{-1}$, whilst Neptune moved towards us with a speed of 9 kms$^{-1}$. This means that  Neptune is receding relative to Uranus, and the expected Doppler red-shift of discrete line emission from Neptune relative to emission from Uranus is 0.22 nm (17 kms$^{-1}$). Therefore, any H$_3^+$ emission present at Neptune will appear 0.22 nm long-ward of the H$_3^+$ emission at Uranus. The spectral data of Neptune presented in this study was shifted by 0.22 nm (14 spectral pixels) towards longer wavelengths to account for this, and all of the following analysis includes this shift. This effectively puts the Neptune spectra in the Uranus rest-frame. Additionally, the iSHELL wavelength dispersion is optimised for use with a 0.375$^{\prime\prime}$ slit at $R\sim70$,000. Since we are using the 1.5$^{\prime\prime}$ slit producing a spectrum at $R\sim17$,500, effectively over-sampling in the spectral domain by a factor of 4, we smooth the spectral data along the wavelength dimension using a rolling boxcar with a width of 4 pixels. We use the wide slit to cover as much as the disk of Neptune as possible, whilst maintaining the ability to guide on Neptune itself.

\begin{figure}
	\centering
	\includegraphics[width=\columnwidth]{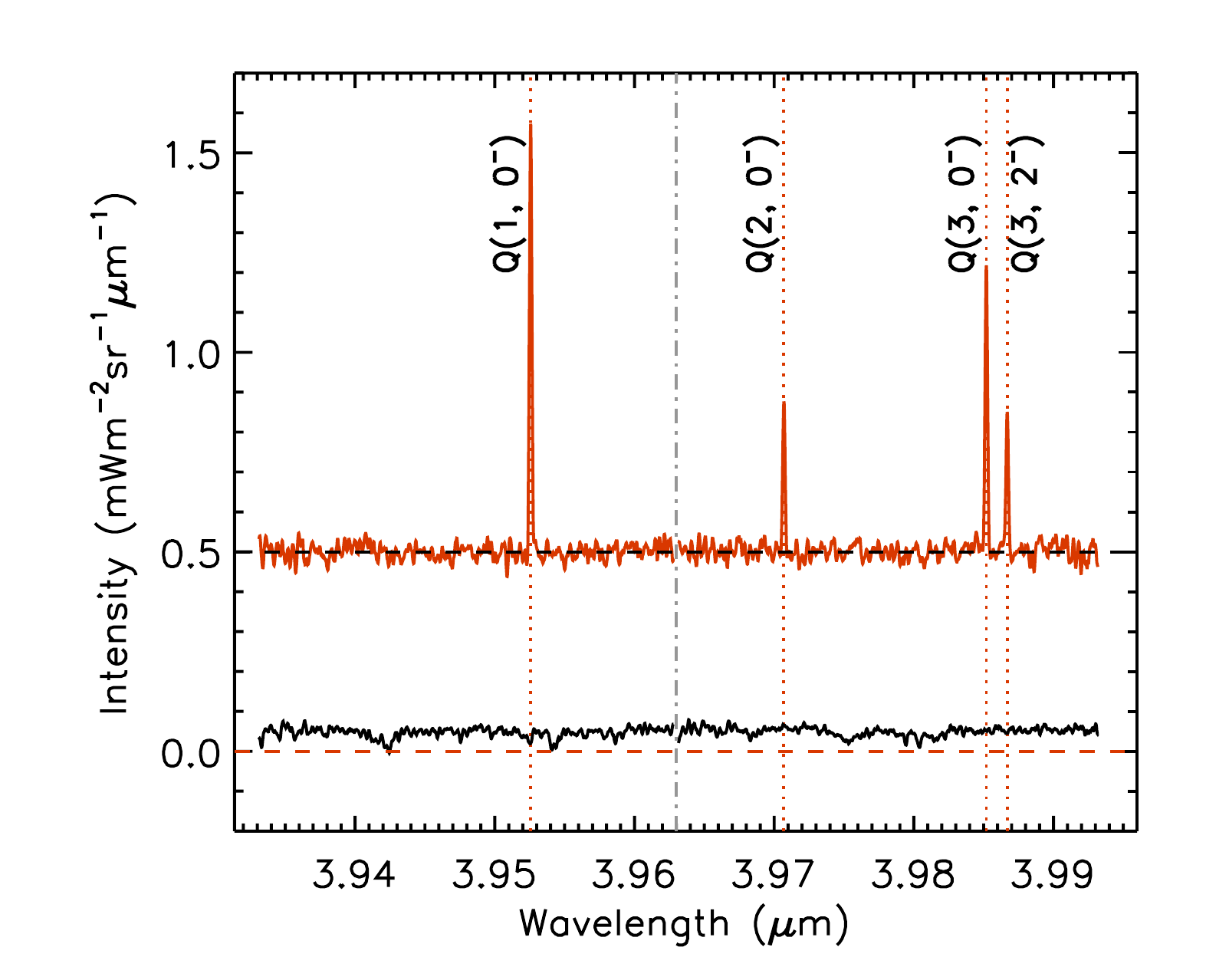}
	\caption{The averaged spectral intensity across the disk of Uranus and Neptune. The Uranus spectra has been shifted by +0.5 mWm$^{-2}$sr$^{-1}$$\mu$m$^{-1}$  for clarity. The dashed lines shows the zero intensity level for each spectrum, the dotted lines indicated the fitted H$_3^+$ line-centres in the Uranus spectrum, and the dot-dashed line shows the boundary between the two spectral orders. The Neptune spectrum is offset from zero, indicating the presence of a continuum emission. The Uranus spectrum shows no evidence of continuum emission and has four discreet H$_3^+$ emission lines.  \label{spectra}}
\end{figure}

\section{Analysis} \label{secanalysis}

The reduced and calibrated individual Neptune observations were stacked and for each pixel, we determined the median value across the 232 A-B pairs, as to produce the average view of the emission from Neptune over the 15.4 hours of observation. The spectral image of the two Neptune iSHELL orders can be seen in Figure \ref{specims}a. The horizontal dimension is spectral and the vertical dimension is spatial, so that the intensity along the slit, across the disk from dawn to dusk, is shown bottom to top. The observations of Neptune in Figure \ref{specims}a shows continuum emission visible along the entire narrow wavelength range, across the entire disk of the planet. This continuum is likely solar reflection from clouds and hazes in the lower atmosphere. 

% 

%Despite the very long integrations, the signal-to-noise of the observed continuum is low ($S/N\sim3$), consistent with a very weak emission. 

The Uranus observations were similarly stacked and medianed, seen in Figure \ref{specims}b, showing clear H$_3^+$ Q(1, 0$^-$) line emission at 3.953 $\mu$m in spectral order 131, and an additional three strong emission lines in the spectral order 130: Q(2, 0$^-$) at 3.971 $\mu$m, Q(3, 0$^-$) at 3.986 $\mu$m, and Q(3, 2$^-$) at 3.994 $\mu$m. The spectral lines are seen across the entire disk, and even slightly above the limbs as a result of a small telescope movements due to uncertainty in the guiding. 

%mixture of an extended upper atmosphere, telluric seeing, and

\begin{figure}
	\centering
	\includegraphics[width=\columnwidth]{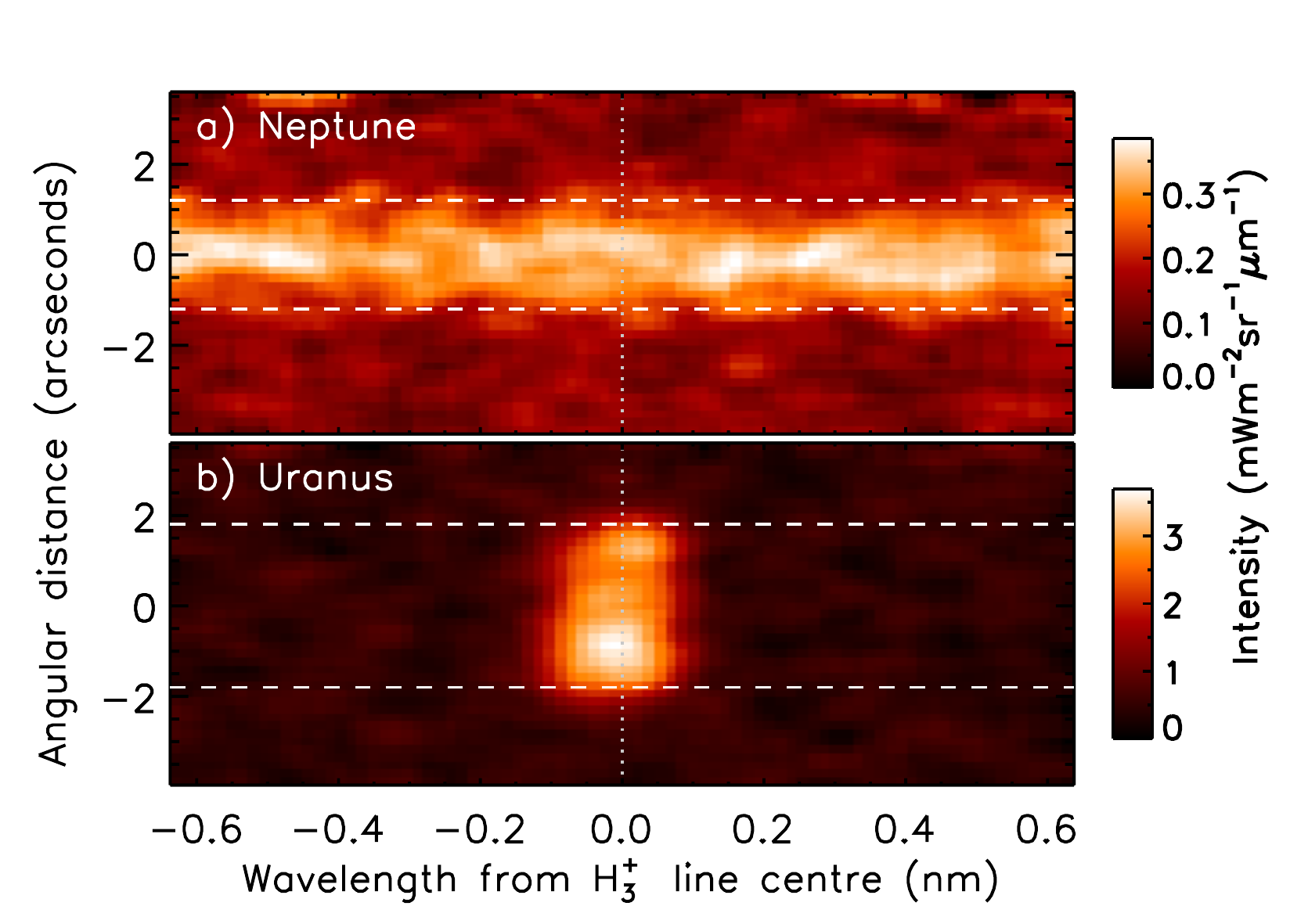}
	\caption{The  four summed H$_3^+$ spectral images of Neptune and Uranus -- see text for details. The horizontal axis shows the wavelength from the H$_3^+$ line centre, indicated by the dotted vertical line, and the vertical axis is the spatial dimension  \label{sspecim}}
\end{figure}

By averaging the spectral intensity in Figure \ref{specims} across the diameter of both Neptune  and Uranus we produce the disk average spectra, shown in Figure \ref{spectra}. The Uranus spectrum has been offset by +0.5 mWm$^{-2}$sr$^{-1}\mu$m$^{-1}$ for clarity. The horizontal dashed lines shows the zero levels for the Neptune and Uranus spectra The grey dot-dashed line shows the boundary between the two iSHELL spectral orders.

The H$_3^+$ line spectrum of Uranus in Figure \ref{spectra}b fits to a temperature of 482$\pm$5 K and a H$_3^+$ column density of 6.8 $\times 10^{15}$ m$^{-2}$. The intergrated H$_3^+$ Q(1, 0$^-$) line intensity is  0.22 $\mu$Wm$^{-2}$sr$^{-1}$. These numbers are similar to those derived by \cite{2011MNRAS.410..641M, 2013Icar..223..741M}, and confirms that the flux calibration that emerges from our iSHELL calibration pipeline is consistent with data from other instruments and facilities. 

No clear and obvious H$_3^+$ line emission in the Neptune spectrum in Figure \ref{spectra} stands out. In order to increase the signal-to-noise, we add the four spectral regions of the Neptune spectral image (Figure \ref{specims}a) that contain H$_3^+$ emission at Uranus. By fitting Gaussians to the four spectral lines in Figure \ref{spectra}, we determine the centre wavelength of each line. These are shown as dotted lines in Figure \ref{spectra}. We then produce a total spectral image of Uranus and Neptune respectively by adding the four individual spectra images centred around each line. These sums of H$_3^+$ Q(1, 0$^-$), Q(2, 0$^-$), Q(3, 0$^-$), and Q(3, 2$^-$) lines are seen in Figure \ref{sspecim}, covering 40 spectral pixels either side of the line centre, indicated by a vertical dotted line. The dashed horizontal lines indicates the spatial extent of each planet.

Figure \ref{sspecim}b shows clear spatial structure in the distribution of intensity across the disk of Uranus. The Uranus observations use a slit rotation generally unsuited for observing the planet, i.e. neither aligned with the equator nor the rotational axis -- see Figure \ref{geometry}. Regardless, analysing the fine details in the Uranus emission remains outside the scope of this study. The sole purpose of the Uranus observations obtained for this study is to accurately determine the expected location of any H$_3^+$ emissions at Neptune. 

The summed spectral images of Neptune in Figure \ref{sspecim}a clearly shows the continuum emission with a peak intensity of about 0.4 mWm$^{-2}$sr$^{-1}\mu$m$^{-1}$. Theres is no obvious intensity enhancement at the expected location of the H$_3^+$ line emission, but it may be difficult to ascertain due to the background continuum. In order to determine if there is emission atop the continuum, we average the emission across the disk of Neptune, and subtract the average value of the continuum emission where H$_3^+$ is not expected to be present. This residual signal is shown in Figure \ref{specsum}a. Figure \ref{specsum}b shows the average intensity of the spectral image in Figure \ref{sspecim}b averaged across the disk of Uranus, showing a strong H$_3^+$ line profile.

Whilst the residual spectral intensity at Neptune in Figure \ref{specsum}a shows a slight enhancement above the zero intensity at the location of the expected H$_3^+$ emission, it is at the level of the surrounding noise. We measure the height of the observed spectral intensity above the mean (zero) level at the expected peak location to be 12$^{+15}_{-10}$ nWm$^{-2}$sr$^{-1}\mu$m$^{-1}$. This is shown as a red dashed line in Figure \ref{specsum}a, with the grey area indicating the sizeable level of uncertainty. Using the H$_3^+$ line-list of \cite{neale_1996} and the partition function of \cite{2010FaDi..147..283M} we can calculate the upper limit of the H$_3^+$ column integrated density by dividing the observed intensity of the four lines added to make Figure \ref{specsum} by the calculated intensity per molecule of the same four lines at a certain temperature. 

%The temperature at the H$_3^+$ density peak using the expected altitude derived by 

\cite{1995Sci...267..648L} derived a H$_3^+$ density peak at 1400 km, at which \cite{1989Sci...246.1459B} measured a temperature of 550 K. Using this temperature, as did \cite{2011MNRAS.410..641M} and \cite{2003AA...403L...7F}, we derive an upper limit of the column integrated H$_3^+$ density of $1.0^{+1.2}_{-0.8}\times10^{13}$ m$^{-2}$. For a temperature of 750 K we derive an upper limit of $0.2^{+0.3}_{-0.16}\times10^{13}$ m$^{-2}$. 

\section{Results \& Discussion} \label{secresults}

Despite 15.4 hours of iSHELL observations of the disk of Neptune we are unable to identify unambiguous H$_3^+$ emissions. The upper limit of the H$_3^+$ column density derived here improves the value of \cite{2011MNRAS.410..641M} by $\sim$30\%. 

\begin{figure}
	\centering
	\includegraphics[width=\columnwidth]{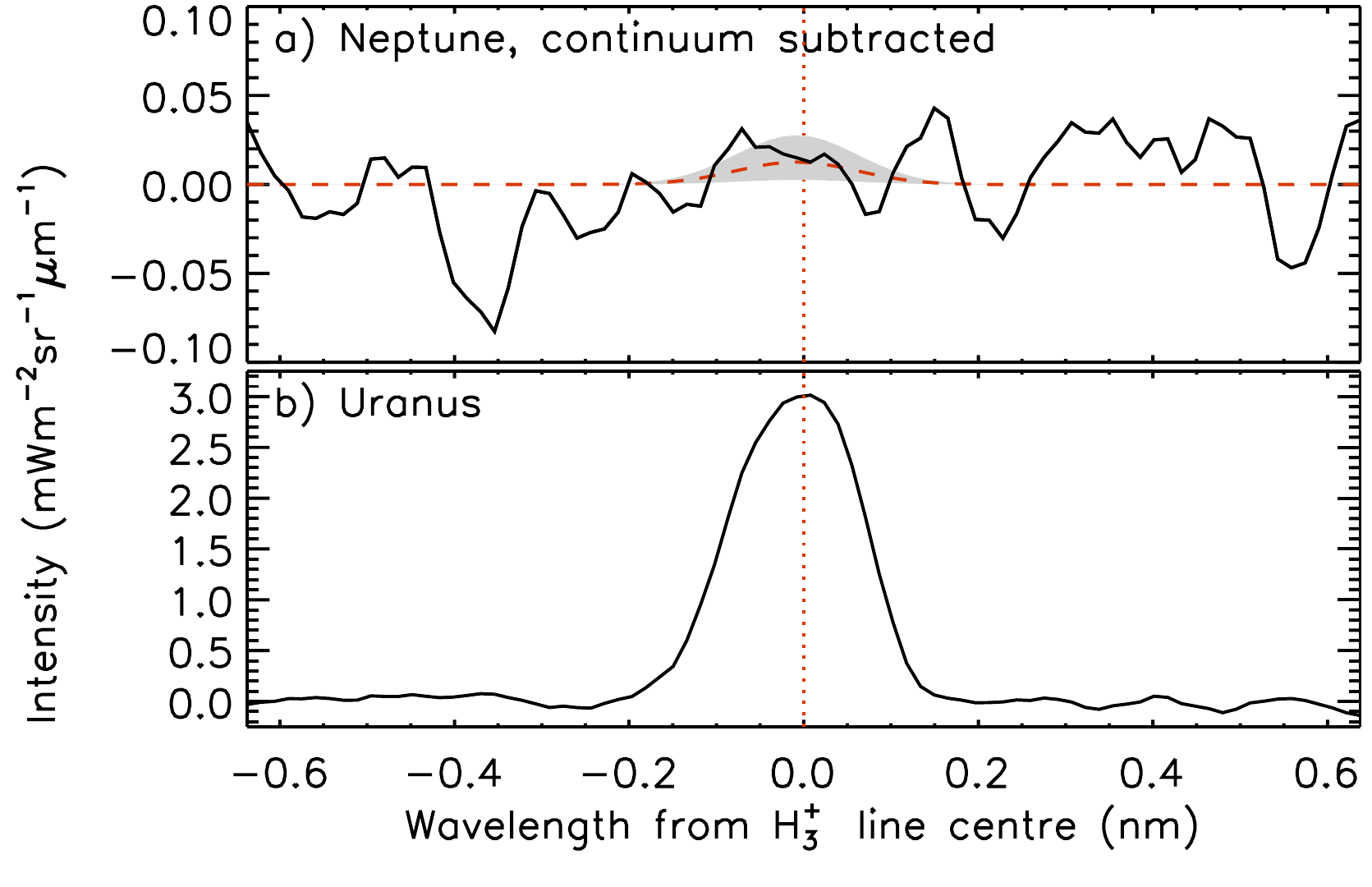}
	\caption{The average spectral intensity contained in the summed spectral images of Neptune and Uranus seen in Figure \ref{sspecim}. The dotted line indicates the H$_3^+$ line centre. a) The average continuum of Neptune has been subtracted off, leaving the residual. The red dashed line indicates the upper limit of the H$_3^+$ line intensity profile, with the grey area indicating the uncertainty range. b) The sum of the four Uranus H$_3^+$ line profiles -- see text for details. \label{specsum}}
\end{figure}
%This improvement is inside the sizeable uncertainty ranges, showing that the two results are consistent. 

The upper limit is about a factor of $\sim$500 times lower than the H$_3^+$ column densities observed at Uranus \citep{2011ApJ...729..134M, 2013Icar..223..741M}, $\sim$100 times lower than at Saturn \citep{2014Icar..229..214O}, and $\sim$1000 times lower than at Jupiter \citep{2002Icar..156..498S, GRL:GRL55734}. One notable difference between Neptune and the other giant planets in our solar system is the presence of CH$_4$, CO$_2$, and H$_2$O at altitudes greater than 500 km above the 1 bar level \citep{2017Icar..297...33M}. The presence of these polyatomic molecules acts as to destroy H$_3^+$ very rapidly via this reaction:
\begin{equation}
H_3^+ + X \rightarrow XH^+ + H_2
\end{equation}
where $X$ is a neutral species heavier  than molecular hydrogen (H$_2$). This recombination would have the effect of vastly shortening the H$_3^+$ lifetime, and whilst the production rate could be comparable to the other giant planets, the very short lifetimes reduce the number of H$_3^+$ molecules to currently un-detectable levels. In this scenario, the amount of H$_3^+$ present in the atmosphere of Neptune would be strongly dependent on the vertical profile of the polyatomic neutrals, which in turn is dependent on the vertical mixing process in the stratosphere \citep{2014Icar..237..211D, 2014Icar..231..146F}; without rigorous vertical mixing, the extended presence of these species into the upper stratosphere and thermopshere cannot be maintained. Therefore, once detected, the column density of H$_3^+$ at Neptune can be used to probe the extent of the vertical mixing in the upper stratosphere. 

% It should be noted that even if Neptune has a strong sink of H$_3^+$ in the upper atmosphere, 

The Keck observations of Neptune by \cite{2011MNRAS.410..641M} showed no continuum emission, which is clearly observed here (see Figure \ref{specims}).  The structure of the near-infrared emission emerging from the lower atmosphere is driven by the presence of clouds \cite[e.g.][]{1998JGR...10323001I}. The continuum emission we observe indicates that the underlying clouds are unable to absorb the incoming sunlight, which is instead scattered back out of the atmosphere. Here, individual nights of observations have slightly different continuum emission profiles across the disk of the planet, suggesting that the clouds are localised in longitude. This emergence of cloud opacity is consistent with the recent development of large storm systems on Neptune observed by \cite{2014Icar..237..211D}. 

\begin{figure*}
	\centering
	\includegraphics[width=7in]{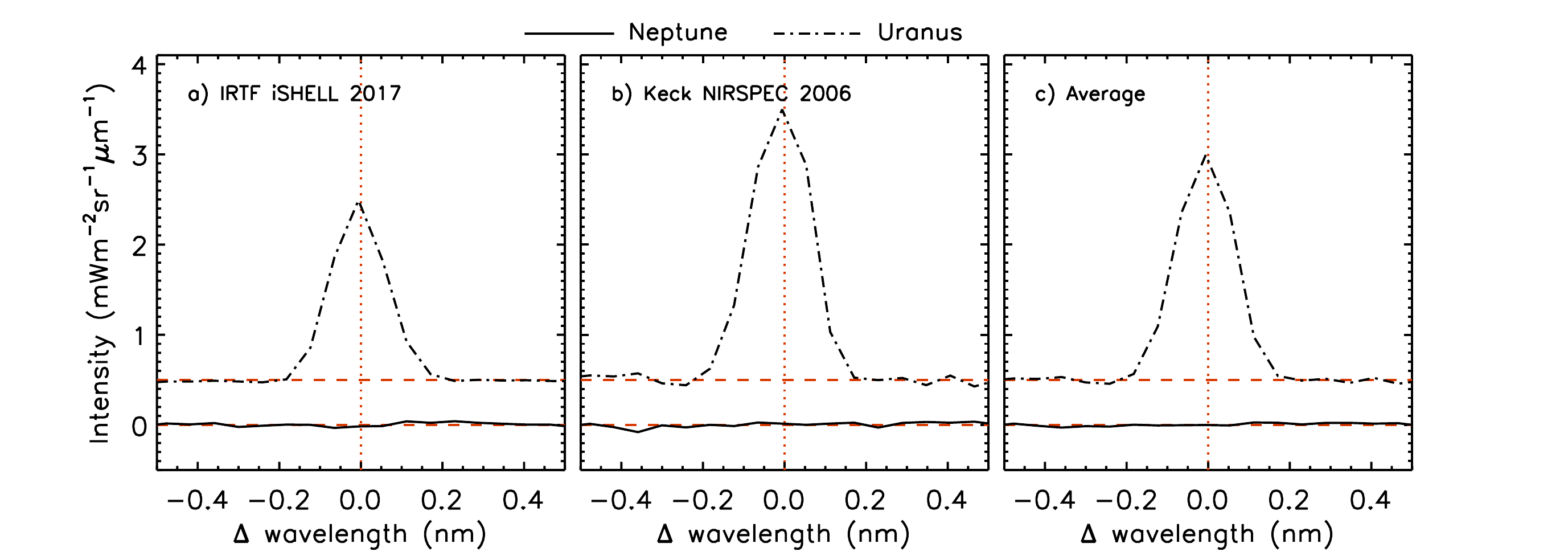}
	\caption{The spectral intensity as a function of distance from the line centre ($\Delta$ wavelength) for Uranus (dot-dashed) and Neptune (solid), comparing the 2017 profiles observed with NASA IRTF iSHELL and the 2006 Keck NIRPEC observations of Melin et al., (2011). Note that the Uranus spectra have been offset by $+0.5$ mWm$^{-2}$sr$^{-1}\mu$m$^{-1}$ for clarity. The horizontal dashed lines indicate the zero intensity level, and the vertical dashed lines indicate the line centre. a) The iSHELL observations are down-sampled to the same wavelength dispersion as the NIRSPEC observations b) The sum of the H$_3^+$ Q(1, 0$^-$), Q(2, 0$^-$), Q(3, 0$^-$), and Q(3, 2$^-$) spectral lines contained in the NIRSPEC data. These are the same lines used in this study. c) The average line line profile of a) and b) for both Uranus and Neptune. From this combined line-profile we derive an upper limit on the H$_3^+$ column density at Neptune of $3.7 \pm 1.4 \times 10^{12}$ m$^{-2}$ at 550 K.\label{ikcomp}}
\end{figure*}

At Jupiter, the presence of storms have been associated with heating in the ionosphere by the breaking of acoustic waves, traveling from the troposphere to the upper atmosphere \citep{2016Natur.536..190O}. With large storms emerging on Neptune, any heating of the ionosphere would have the effect of increasing the intensity of any H$_3^+$ emission; the intensity is driven exponentially by  temperature, and linearly by the H$_3^+$ density \citep{2014MNRAS.438.1611M}. Despite this potential source of increased H$_3^+$ intensities, we are unable to detect H$_3^+$ at Neptune.

% --  --  --  --  --  --  --  --  --  --  --  --  --  --  --  --  --  --

The observations presented here span 15.4 h over 4 days, with a spectrograph slit that covers a significant fraction of the disk of Neptune (see Figure \ref{geometry}). Since the rotation period is 16.1 h, the observations effectively provide a longitude averaged view, removing any short-term variability driven by changes in viewing geometry. The detection attempt of \cite{2011MNRAS.410..641M}, using Keck observations from 2006, is separated from the 2017 observations presented here by almost exactly one solar cycle. Therefore, we may expect the levels of solar EUV flux to be similar, resulting in similar levels of H$_3^+$ produced via solar ionisation. We do not yet know what the `normal' baseline levels of H$_3^+$ emissions are at Neptune, and a detection would enable us to characterise the upper atmosphere and aurora, and associated variability, for the first time. A 30\% reduction in the upper limit of the H$_3^+$ intensity from the previous estimate of \cite{2011MNRAS.410..641M} means that we are moving further below the model predictions of \cite{1995Sci...267..648L}, with the model over-estimating the H$_3^+$ density by at least a factor of 5. Consequently, there is clearly potential scope for developing new modelling efforts at Neptune.

The iSHELL slit-width used in this study (1.5$^{\prime\prime}$) produces a spectrum at a similar spectral resolution to that of Keck NIRSPEC, used by \cite{2011MNRAS.410..641M}. Assuming that similar solar conditions prevailed, it can be instructive to average the two datasets to investigate if the upper limit can be improved upon, with the understanding that such an exercise can be fraught with complications. In Figure \ref{ikcomp}a we have down-sampled the iSHELL observations of Figure \ref{specsum} (0.016 nm/pixel) to the same wavelength dispersion as the NIRSPEC observations (0.060 nm/pixel). Figure \ref{ikcomp}b shows the sum of the H$_3^+$ Q(1, 0$^-$), Q(2, 0$^-$), Q(3, 0$^-$), and Q(3, 2$^-$) lines of Uranus and Neptune contained within the Keck dataset of \cite{2011MNRAS.410..641M}. The difference in the spectral intensity of the Uranus H$_3^+$ line profiles in Figure \ref{ikcomp}a and \ref{ikcomp}b is likely driven by differences in both temperature and H$_3^+$ density between 2006 and 2017: we derive a temperature for the 2017 data of 482$\pm$5 K, whereas the temperature derived from the 2006 Keck data was 608$\pm$12 K \citep{2011ApJ...729..134M}. Figure \ref{ikcomp} shows the average Uranus and Neptune profiles, and we derive an upper limit on the H$_3^+$ density of $0.37 \pm 0.14 \times 10^{13}$ m$^{-2}$ (at 550 K), a factor of 2.7 lower than derived from the iSHELL data alone. Therefore, the combination of the two datasets suggests that the model of \cite{1995Sci...267..648L} overestimates the H$_3^+$ density by at least a factor of 13.

% Uranus was  observed to be 608$\pm$12 K in 2006, whilst we derive a temperature of 482$\pm$5 K. This difference corresponds to a density difference of 

At Uranus, the arrival of solar wind compressions have been shown to increase the ultraviolet auroral emissions \citep{2012GeoRL..39.7105L}. An increased particle precipitation flux could also lead to an increase in the production of H$_3^+$. One potential strategy for future H$_3^+$ detection attempts at Neptune is to time the iSHELL  observations to coincide with the arrival of such compressions, potentially producing more intense H$_3^+$ auroral emission. However, there remains significant uncertainty in modelling the propagation of these solar wind features observed at Earth out to the orbit of Neptune at 39 AU. 

% --  --  --  --  --  --  --  --  --  --  --  --  --  --  --  --  --  --

The instrument suite on-board the James Webb Space Telescope (JWST), due to be launched in 2018, includes a near-infrared spectrograph \citep[NIRSpec,][]{2007SPIE.6692E..0MB} capable of observing H$_3^+$ emissions from the giant planets \citep{2016PASP..128a8005N}. Detecting H$_3^+$ emissions from Neptune would open many significant avenues of scientific inquiry, and JWST emerges as the next step in this particular journey. Our present study provides limits on the expected intensity of the H$_3^+$ emissions.

\section{Conclusions}\label{secconc}

We have analysed NASA IRTF iSHELL observations of Neptune obtained over four nights in August 2017 in an attempt to detect H$_3^+$ emission from the planet. Despite 15.4 hours of integrating across the disk of the planet, covering all longitudes, we are unable to make a positive detection. Instead, we derive an upper limit of the H$_3^+$ column density of $1.0^{+1.2}_{-0.8}\times10^{13}$ m$^{-2}$ for a temperature of 550 K. 

The upper limit for the H$_3^+$ column density is about 500 times less than is observed at Uranus. This difference could be driven by the fact that Neptune has strong vertical mixing that moves CH$_4$, CO$_2$, and H$_2$O from the upper stratosphere into the thermosphere, which acts as to destroy H$_3^+$. This shortens H$_3^+$ lifetimes, and reduces the H$_3^+$ density to currently un-detectable levels.

\section*{Acknowledgements}\label{seckack}
This work was supported by the UK Science and Technology Facilities Council (STFC) Grant ST/N000749/1 for HM and TSS. LNF was supported by a Royal Society Research Fellowship at the University of Leicester. REJ and PTD were supported by STFC studentships. Support for  JO'D. comes from an appointment to the NASA Postdoctoral Program at Goddard Space Flight Center, administered by Universities Space Research Association under contract with NASA. LM was supported by NASA under Grant NNX17AF14G issued through the SSO Planetary Astronomy Program. HM, REJ, and TSS are Visiting Astronomers at the Infrared Telescope Facility, which is operated by the University of Hawaii under contract NNH14CK55B with the National Aeronautics and Space Administration. We would like to express our gratitude to Steve Miller for helpful comments and suggestions.

\bibliographystyle{plainnat}

\bibliography{neptune}

\begin{thebibliography}{31}
\providecommand{\natexlab}[1]{#1}
\providecommand{\url}[1]{\texttt{#1}}
\expandafter\ifx\csname urlstyle\endcsname\relax
  \providecommand{\doi}[1]{doi: #1}\else
  \providecommand{\doi}{doi: \begingroup \urlstyle{rm}\Url}\fi

\bibitem[{Bagnasco} et~al.(2007){Bagnasco}, {Kolm}, {Ferruit}, {Honnen},
  {Koehler}, {Lemke}, {Maschmann}, {Melf}, {Noyer}, {Rumler}, {Salvignol},
  {Strada}, and {Te Plate}]{2007SPIE.6692E..0MB}
G.~{Bagnasco}, M.~{Kolm}, P.~{Ferruit}, K.~{Honnen}, J.~{Koehler}, R.~{Lemke},
  M.~{Maschmann}, M.~{Melf}, G.~{Noyer}, P.~{Rumler}, J.-C. {Salvignol},
  P.~{Strada}, and M.~{Te Plate}.
\newblock {Overview of the near-infrared spectrograph (NIRSpec) instrument
  on-board the James Webb Space Telescope (JWST)}.
\newblock In \emph{Cryogenic Optical Systems and Instruments XII}, volume 6692
  of \emph{\procspie}, page 66920M, September 2007.
\newblock \doi{10.1117/12.735602}.

\bibitem[{Broadfoot} et~al.(1989){Broadfoot}, {Atreya}, {Bertaux}, {Blamont},
  {Dessler}, {Donahue}, {Forrester}, {Hall}, {Herbert}, {Holberg}, {Hunten},
  {Krasnopolsky}, {Linick}, {Lunine}, {Mcconnell}, {Moos}, {Sandel},
  {Schneider}, {Shemansky}, {Smith}, {Strobel}, and
  {Yelle}]{1989Sci...246.1459B}
A.~L. {Broadfoot}, S.~K. {Atreya}, J.~L. {Bertaux}, J.~E. {Blamont}, A.~J.
  {Dessler}, T.~M. {Donahue}, W.~T. {Forrester}, D.~T. {Hall}, F.~{Herbert},
  J.~B. {Holberg}, D.~M. {Hunten}, V.~A. {Krasnopolsky}, S.~{Linick}, J.~I.
  {Lunine}, J.~C. {Mcconnell}, H.~W. {Moos}, B.~R. {Sandel}, N.~M. {Schneider},
  D.~E. {Shemansky}, G.~R. {Smith}, D.~F. {Strobel}, and R.~V. {Yelle}.
\newblock {Ultraviolet spectrometer observations of Neptune and Triton}.
\newblock \emph{Science}, 246:\penalty0 1459--1466, December 1989.
\newblock \doi{10.1126/science.246.4936.1459}.

\bibitem[{de Pater} et~al.(2014){de Pater}, {Fletcher}, {Luszcz-Cook},
  {DeBoer}, {Butler}, {Hammel}, {Sitko}, {Orton}, and
  {Marcus}]{2014Icar..237..211D}
I.~{de Pater}, L.~N. {Fletcher}, S.~{Luszcz-Cook}, D.~{DeBoer}, B.~{Butler},
  H.~B. {Hammel}, M.~L. {Sitko}, G.~{Orton}, and P.~S. {Marcus}.
\newblock {Neptune's global circulation deduced from multi-wavelength
  observations}.
\newblock \emph{\icarus}, 237:\penalty0 211--238, July 2014.
\newblock \doi{10.1016/j.icarus.2014.02.030}.

\bibitem[Dinelli et~al.(2017)Dinelli, Fabiano, Adriani, Altieri, Moriconi,
  Mura, Sindoni, Filacchione, Tosi, Migliorini, Grassi, Piccioni, Noschese,
  Cicchetti, Bolton, Connerney, Atreya, Bagenal, Gladstone, Hansen, Kurth,
  Levin, Mauk, McComas, G{\`e}rard, Turrini, Stefani, Amoroso, and
  Olivieri]{GRL:GRL55734}
B.~M. Dinelli, F.~Fabiano, A.~Adriani, F.~Altieri, M.~L. Moriconi, A.~Mura,
  G.~Sindoni, G.~Filacchione, F.~Tosi, A.~Migliorini, D.~Grassi, G.~Piccioni,
  R.~Noschese, A.~Cicchetti, S.~J. Bolton, J.~E.~P. Connerney, S.~K. Atreya,
  F.~Bagenal, G.~R. Gladstone, C.~J. Hansen, W.~S. Kurth, S.~M. Levin, B.~H.
  Mauk, D.~J. McComas, J.-C. G{\`e}rard, D.~Turrini, S.~Stefani, M.~Amoroso,
  and A.~Olivieri.
\newblock Preliminary jiram results from juno polar observations: 1.
  methodology and analysis applied to the jovian northern polar region.
\newblock \emph{Geophysical Research Letters}, 44\penalty0 (10):\penalty0
  4625--4632, 2017.
\newblock ISSN 1944-8007.
\newblock \doi{10.1002/2017GL072929}.
\newblock URL \url{http://dx.doi.org/10.1002/2017GL072929}.
\newblock 2017GL072929.

\bibitem[{Encrenaz} et~al.(2000){Encrenaz}, {Schulz}, {Drossart}, {Lellouch},
  {Feuchtgruber}, and {Atreya}]{2000AA...358L..83E}
T.~{Encrenaz}, B.~{Schulz}, P.~{Drossart}, E.~{Lellouch}, H.~{Feuchtgruber},
  and S.~K. {Atreya}.
\newblock {The ISO spectra of Uranus and Neptune between 2.5 and 4.2 mu m:
  constraints on albedos and H$_3^{+}$}.
\newblock \emph{{Astronomy and Astrophysics}}, 358:\penalty0 L83--L87, June
  2000.

\bibitem[{Feuchtgruber} and {Encrenaz}(2003)]{2003AA...403L...7F}
H.~{Feuchtgruber} and T.~{Encrenaz}.
\newblock {The infrared spectrum of Neptune at 3.5-4.1 microns: Search for
  H$_{3}^{+}$ and evidence for recent meteorological variations}.
\newblock \emph{{Astronomy and Astrophysics}}, 403:\penalty0 L7--L10, May 2003.
\newblock \doi{10.1051/0004-6361:20030414}.

\bibitem[{Fletcher} et~al.(2014){Fletcher}, {de Pater}, {Orton}, {Hammel},
  {Sitko}, and {Irwin}]{2014Icar..231..146F}
L.~N. {Fletcher}, I.~{de Pater}, G.~S. {Orton}, H.~B. {Hammel}, M.~L. {Sitko},
  and P.~G.~J. {Irwin}.
\newblock {Neptune at summer solstice: Zonal mean temperatures from
  ground-based observations, 2003-2007}.
\newblock \emph{\icarus}, 231:\penalty0 146--167, March 2014.
\newblock \doi{10.1016/j.icarus.2013.11.035}.

\bibitem[{Herbert}(2009)]{2009JGRA..11411206H}
F.~{Herbert}.
\newblock {Aurora and magnetic field of Uranus}.
\newblock \emph{Journal of Geophysical Research (Space Physics)}, 114\penalty0
  (13):\penalty0 11206--+, November 2009.
\newblock \doi{10.1029/2009JA014394}.

\bibitem[{Irwin} et~al.(1998){Irwin}, {Weir}, {Smith}, {Taylor}, {Lambert},
  {Calcutt}, {Cameron-Smith}, {Carlson}, {Baines}, {Orton}, {Drossart},
  {Encrenaz}, and {Roos-Serote}]{1998JGR...10323001I}
P.~G.~J. {Irwin}, A.~L. {Weir}, S.~E. {Smith}, F.~W. {Taylor}, A.~L. {Lambert},
  S.~B. {Calcutt}, P.~J. {Cameron-Smith}, R.~W. {Carlson}, K.~{Baines}, G.~S.
  {Orton}, P.~{Drossart}, T.~{Encrenaz}, and M.~{Roos-Serote}.
\newblock {Cloud structure and atmospheric composition of Jupiter retrieved
  from Galileo near-infrared mapping spectrometer real-time spectra}.
\newblock \emph{\jgr}, 103:\penalty0 23001--23022, September 1998.
\newblock \doi{10.1029/98JE00948}.

\bibitem[{Lamy} et~al.(2012){Lamy}, {Prang{\'e}}, {Hansen}, {Clarke}, {Zarka},
  {Cecconi}, {Aboudarham}, {Andr{\'e}}, {Branduardi-Raymont}, {Gladstone},
  {Barth{\'e}l{\'e}my}, {Achilleos}, {Guio}, {Dougherty}, {Melin}, {Cowley},
  {Stallard}, {Nichols}, and {Ballester}]{2012GeoRL..39.7105L}
L.~{Lamy}, R.~{Prang{\'e}}, K.~C. {Hansen}, J.~T. {Clarke}, P.~{Zarka},
  B.~{Cecconi}, J.~{Aboudarham}, N.~{Andr{\'e}}, G.~{Branduardi-Raymont},
  R.~{Gladstone}, M.~{Barth{\'e}l{\'e}my}, N.~{Achilleos}, P.~{Guio}, M.~K.
  {Dougherty}, H.~{Melin}, S.~W.~H. {Cowley}, T.~S. {Stallard}, J.~D.
  {Nichols}, and G.~{Ballester}.
\newblock {Earth-based detection of Uranus' aurorae}.
\newblock \emph{\grl}, 39:\penalty0 L07105, April 2012.
\newblock \doi{10.1029/2012GL051312}.

\bibitem[{Lyons}(1995)]{1995Sci...267..648L}
J.~R. {Lyons}.
\newblock {Metal Ions in the Atmosphere of Neptune}.
\newblock \emph{Science}, 267:\penalty0 648--651, February 1995.
\newblock \doi{10.1126/science.7839139}.

\bibitem[{Masters}(2015)]{2015JGRA..120..479M}
A.~{Masters}.
\newblock {Magnetic reconnection at Neptune's magnetopause}.
\newblock \emph{Journal of Geophysical Research (Space Physics)}, 120:\penalty0
  479--493, January 2015.
\newblock \doi{10.1002/2014JA020744}.

\bibitem[{McLean} et~al.(1998){McLean}, {Becklin}, {Bendiksen}, {Brims},
  {Canfield}, {Figer}, {Graham}, {Hare}, {Lacayanga}, {Larkin}, {Larson},
  {Levenson}, {Magnone}, {Teplitz}, and {Wong}]{1998SPIE.3354..566M}
I.~S. {McLean}, E.~E. {Becklin}, O.~{Bendiksen}, G.~{Brims}, J.~{Canfield},
  D.~F. {Figer}, J.~R. {Graham}, J.~{Hare}, F.~{Lacayanga}, J.~E. {Larkin},
  S.~B. {Larson}, N.~{Levenson}, N.~{Magnone}, H.~{Teplitz}, and W.~{Wong}.
\newblock {Design and development of NIRSPEC: a near-infrared echelle
  spectrograph for the Keck II telescope}.
\newblock In {A.~M.~Fowler}, editor, \emph{Society of Photo-Optical
  Instrumentation Engineers (SPIE) Conference Series}, volume 3354 of
  \emph{Presented at the Society of Photo-Optical Instrumentation Engineers
  (SPIE) Conference}, pages 566--578, August 1998.

\bibitem[{Melin} et~al.(2011{\natexlab{a}}){Melin}, {Stallard}, {Miller},
  {Lystrup}, {Trafton}, {Booth}, and {Rivers}]{2011MNRAS.410..641M}
H.~{Melin}, T.~{Stallard}, S.~{Miller}, M.~B. {Lystrup}, L.~M. {Trafton}, T.~C.
  {Booth}, and C.~{Rivers}.
\newblock {New limits on H$^{+}$$_{3}$ abundance on Neptune using Keck
  NIRSPEC}.
\newblock \emph{\mnras}, 410:\penalty0 641--644, January 2011{\natexlab{a}}.
\newblock \doi{10.1111/j.1365-2966.2010.17468.x}.

\bibitem[{Melin} et~al.(2011{\natexlab{b}}){Melin}, {Stallard}, {Miller},
  {Trafton}, {Encrenaz}, and {Geballe}]{2011ApJ...729..134M}
H.~{Melin}, T.~{Stallard}, S.~{Miller}, L.~M. {Trafton}, T.~{Encrenaz}, and
  T.~R. {Geballe}.
\newblock {Seasonal Variability in the Ionosphere of Uranus}.
\newblock \emph{\apj}, 729:\penalty0 134, March 2011{\natexlab{b}}.
\newblock \doi{10.1088/0004-637X/729/2/134}.

\bibitem[{Melin} et~al.(2013){Melin}, {Stallard}, {Miller}, {Geballe},
  {Trafton}, and {O'Donoghue}]{2013Icar..223..741M}
H.~{Melin}, T.~S. {Stallard}, S.~{Miller}, T.~R. {Geballe}, L.~M. {Trafton},
  and J.~{O'Donoghue}.
\newblock {Post-equinoctial observations of the ionosphere of Uranus}.
\newblock \emph{\icarus}, 223:\penalty0 741--748, April 2013.
\newblock \doi{10.1016/j.icarus.2013.01.012}.

\bibitem[{Melin} et~al.(2014){Melin}, {Stallard}, {O'Donoghue}, {Badman},
  {Miller}, and {Blake}]{2014MNRAS.438.1611M}
H.~{Melin}, T.~S. {Stallard}, J.~{O'Donoghue}, S.~V. {Badman}, S.~{Miller}, and
  J.~S.~D. {Blake}.
\newblock {On the anticorrelation between H$_3^+$ temperature and density in
  giant planet ionospheres}.
\newblock \emph{\mnras}, 438:\penalty0 1611--1617, February 2014.
\newblock \doi{10.1093/mnras/stt2299}.

\bibitem[{Miller}(2010)]{miller_2010}
S.~{Miller}.
\newblock {H$_3^+$ in the thermospheres of the giant planets}.
\newblock \emph{{Some Journal}}, 2010.

\bibitem[{Miller} et~al.(2010){Miller}, {Stallard}, {Melin}, and
  {Tennyson}]{2010FaDi..147..283M}
S.~{Miller}, T.~{Stallard}, H.~{Melin}, and J.~{Tennyson}.
\newblock {H$_3^+$ cooling in planetary atmospheres}.
\newblock \emph{Faraday Discussions}, 147:\penalty0 283--+, 2010.
\newblock \doi{10.1039/c004152c}.

\bibitem[{Moses} and {Poppe}(2017)]{2017Icar..297...33M}
J.~I. {Moses} and A.~R. {Poppe}.
\newblock {Dust ablation on the giant planets: Consequences for stratospheric
  photochemistry}.
\newblock \emph{\icarus}, 297:\penalty0 33--58, November 2017.
\newblock \doi{10.1016/j.icarus.2017.06.002}.

\bibitem[{Neale} et~al.(1996){Neale}, {Miller}, and {Tennyson}]{neale_1996}
L.~{Neale}, S.~{Miller}, and J.~{Tennyson}.
\newblock {Spectroscopic Properties of the H$_3^+$ Molecule: A New Calculated
  Line List}.
\newblock \emph{The Astrophysical Journal}, 464:\penalty0 516--+, 1996.

\bibitem[{Ness} et~al.(1989){Ness}, {Acuna}, {Burlaga}, {Connerney}, and
  {Lepping}]{1989Sci...246.1473N}
N.~F. {Ness}, M.~H. {Acuna}, L.~F. {Burlaga}, J.~E.~P. {Connerney}, and R.~P.
  {Lepping}.
\newblock {Magnetic fields at Neptune}.
\newblock \emph{Science}, 246:\penalty0 1473--1478, December 1989.
\newblock \doi{10.1126/science.246.4936.1473}.

\bibitem[{Norwood} et~al.(2016){Norwood}, {Moses}, {Fletcher}, {Orton},
  {Irwin}, {Atreya}, {Rages}, {Cavali{\'e}}, {S{\'a}nchez-Lavega}, {Hueso}, and
  {Chanover}]{2016PASP..128a8005N}
J.~{Norwood}, J.~{Moses}, L.~N. {Fletcher}, G.~{Orton}, P.~G.~J. {Irwin},
  S.~{Atreya}, K.~{Rages}, T.~{Cavali{\'e}}, A.~{S{\'a}nchez-Lavega},
  R.~{Hueso}, and N.~{Chanover}.
\newblock {Giant Planet Observations with the James Webb Space Telescope}.
\newblock \emph{\pasp}, 128\penalty0 (1):\penalty0 018005, January 2016.
\newblock \doi{10.1088/1538-3873/128/959/018005}.

\bibitem[{O'Donoghue} et~al.(2014){O'Donoghue}, {Stallard}, {Melin}, {Cowley},
  {Badman}, {Moore}, {Miller}, {Tao}, {Baines}, and
  {Blake}]{2014Icar..229..214O}
J.~{O'Donoghue}, T.~S. {Stallard}, H.~{Melin}, S.~W.~H. {Cowley}, S.~V.
  {Badman}, L.~{Moore}, S.~{Miller}, C.~{Tao}, K.~H. {Baines}, and J.~S.~D.
  {Blake}.
\newblock {Conjugate observations of Saturn's northern and southern H$_3^+$
  aurorae}.
\newblock \emph{\icarus}, 229:\penalty0 214--220, February 2014.
\newblock \doi{10.1016/j.icarus.2013.11.009}.

\bibitem[{O'Donoghue} et~al.(2016){O'Donoghue}, {Moore}, {Stallard}, and
  {Melin}]{2016Natur.536..190O}
J.~{O'Donoghue}, L.~{Moore}, T.~S. {Stallard}, and H.~{Melin}.
\newblock {Heating of Jupiter's upper atmosphere above the Great Red Spot}.
\newblock \emph{\nat}, 536:\penalty0 190--192, August 2016.
\newblock \doi{10.1038/nature18940}.

\bibitem[{Rayner} et~al.(2016){Rayner}, {Tokunaga}, {Jaffe}, {Bonnet}, {Ching},
  {Connelley}, {Kokubun}, {Lockhart}, and {Warmbier}]{2016SPIE.9908E..84R}
J.~{Rayner}, A.~{Tokunaga}, D.~{Jaffe}, M.~{Bonnet}, G.~{Ching},
  M.~{Connelley}, D.~{Kokubun}, C.~{Lockhart}, and E.~{Warmbier}.
\newblock {iSHELL: a construction, assembly and testing}.
\newblock In \emph{Ground-based and Airborne Instrumentation for Astronomy VI},
  volume 9908 of \emph{\procspie}, page 990884, August 2016.
\newblock \doi{10.1117/12.2232064}.

\bibitem[{Sandel} et~al.(1990){Sandel}, {Herbert}, {Dessler}, and
  {Hill}]{1990GeoRL..17.1693S}
B.~R. {Sandel}, F.~{Herbert}, A.~J. {Dessler}, and T.~W. {Hill}.
\newblock {Aurora and airglow on the night side of Neptune}.
\newblock \emph{{Geophysical Research Letters}}, 17:\penalty0 1693--1696,
  September 1990.
\newblock \doi{10.1029/GL017i010p01693}.

\bibitem[{Stallard} et~al.(2002){Stallard}, {Miller}, {Millward}, and
  {Joseph}]{2002Icar..156..498S}
T.~{Stallard}, S.~{Miller}, G.~{Millward}, and R.~D. {Joseph}.
\newblock {On the Dynamics of the Jovian Ionosphere and Thermosphere. II. The
  Measurement of H $_{3}$$^{+}$ Vibrational Temperature, Column Density, and
  Total Emission}.
\newblock \emph{\icarus}, 156:\penalty0 498--514, April 2002.
\newblock \doi{10.1006/icar.2001.6793}.

\bibitem[{Trafton} et~al.(1993){Trafton}, {Geballe}, {Miller}, {Tennyson}, and
  {Ballester}]{1993ApJ...405..761T}
L.~M. {Trafton}, T.~R. {Geballe}, S.~{Miller}, J.~{Tennyson}, and G.~E.
  {Ballester}.
\newblock {Detection of H$_3^+$ from Uranus}.
\newblock \emph{The Astrophysical Journal}, 405:\penalty0 761--766, March 1993.
\newblock \doi{10.1086/172404}.

\bibitem[{Warwick} et~al.(1989){Warwick}, {Evans}, {Peltzer}, {Peltzer},
  {Romig}, {Sawyer}, {Riddle}, {Schweitzer}, {Desch}, and
  {Kaiser}]{1989Sci...246.1498W}
J.~W. {Warwick}, D.~R. {Evans}, G.~R. {Peltzer}, R.~G. {Peltzer}, J.~H.
  {Romig}, C.~B. {Sawyer}, A.~C. {Riddle}, A.~E. {Schweitzer}, M.~D. {Desch},
  and M.~L. {Kaiser}.
\newblock {Voyager planetary radio astronomy at Neptune}.
\newblock \emph{Science}, 246:\penalty0 1498--1501, December 1989.
\newblock \doi{10.1126/science.246.4936.1498}.

\bibitem[{Yelle} and {Miller}(2004)]{2004jpsm.book..185Y}
R.~V. {Yelle} and S.~{Miller}.
\newblock \emph{{Jupiter's thermosphere and ionosphere}}, pages 185--218.
\newblock 2004.

\end{thebibliography}

% Don't change these lines
\bsp	% typesetting comment
\label{lastpage}
\end{document}